\documentstyle{elsart}
\begin{document}

\begin{frontmatter}
\title{On the Umemura Polynomials for the Painlev\'e III equation}

\author{Kenji Kajiwara and Tetsu Masuda}

\address{Department of Electrical Engineering, \\
         Doshisha University, Kyotanabe, Kyoto 610-0321, Japan}

\begin{abstract}
A determinant expression for the rational solutions of the Painlev\'e III (P$_{\rm III}$) 
equation whose entries are the Laguerre polynomials is given.
Degeneration of this determinant expression to that for the rational solutions 
of P$_{\rm II}$ is discussed by applying the coalescence procedure. 
\end{abstract}
\end{frontmatter}
\maketitle

\section{Introduction}

The role of Painlev\'e equations in nonlinear world is well recognized. 
Originally Painlev\'e derived these equations to find new transcendental functions, 
and irreducibility of the solutions has been established 
in general by Umemura~\cite{Umemura:100years}. 
However, classical solutions of the Painlev\'e equations have attractive and mysterious properties.
Recently, such properties of rational solutions have been studied extensively. 
Vorob'ev and Yablonskii have shown that the rational solutions of P$_{\rm II}$ 
are expressed by log derivative of some polynomials which are now called 
the Yablonskii-Vorob'ev polynomials~\cite{YV}. 
Okamoto have shown that the rational solutions of P$_{\rm IV}$ have the same property. 
Moreover, he also noticed that they are located on special points in the parameter space 
from the point of symmetry~\cite{Okamoto}. 
Namely, they are in the barycenter of Weyl chamber associated 
with the Affine Weyl group of $A_2^{(1)}$ type which is the transformation group of P$_{\rm IV}$. 
Umemura has shown that there exist special polynomials 
for P$_{\rm III}$, P$_{\rm V}$ and P$_{\rm VI}$, 
which admit the similar properties to those of P$_{\rm II}$ and P$_{\rm IV}$~\cite{Umemura1}.
Those polynomials are called the Umemura polynomials. 
Moreover, it is reported that they have mysterious combinatorial property~\cite{Umemura2}. 

One aspect of such polynomials associated with P$_{\rm II}$\cite{p2:rational}, 
P$_{\rm IV}$\cite{p4:rational,NY:P4} and P$_{\rm V}$\cite{NY} is that 
they are expressed as special cases of the Schur polynomials, 
which arise as the $\tau$-function of the celebrated KP hierarchy as well 
as the character polynomials of symmetric groups. 
Note that the symmetries of these three equations are all $A$ type, 
namely, $A_l^{(1)}$, $l=1,2,3$ for P$_{\rm II}$, P$_{\rm IV}$ and P$_{\rm V}$, respectively. 

In this letter, we consider the Umemura polynomials for P$_{\rm III}$ 
whose symmetry is $B_2^{(1)}$. 
We show that they are expressed as the special case of 2-reduced Schur polynomials. 
Coalescence limit to the Yablonskii-Vorob'ev polynomials is also discussed.

\section{A determinant expression of the Umemura polynomials for P$_{\rm III}$}
The Umemura polynomials for P$_{\rm III}$, 
\begin{equation}
\frac{d^2w}{dr^2}=\frac{1}{w}\left(\frac{dw}{dr}\right)^2-\frac{1}{r}\frac{dw}{dr}
                  +\frac{\alpha w^2+\beta}{r}+4w^3-\frac{4}{w},   \label{P3}
\end{equation}
are defined as follows\cite{Umemura1}. Let $S_N=S_N(u,v)$ be a sequence of polynomials in $u$ defined 
through the recurrence relation of Toda type, 
\begin{eqnarray}
S_{N+1}S_{N-1}=-4u^4&&\left[\frac{d^2S_N}{du^2}S_N-\left(\frac{dS_N}{du}\right)^2\right]
\nonumber\\
&& -4u^3\frac{dS_N}{du}S_N+\left[1+(2v+1)u\right]S_N^2, 
\end{eqnarray}
with initial conditions $S_0=S_1=1$. If we set $u=1/(4r)$ and $S_N(u,v)=T_N(r,v)$, 
we find that $T_N$ satisfy the following recursion relation, 
\begin{equation}
\hspace*{-20pt}
4r T_{N+1}T_{N-1}=-r\left[\frac{d^2T_N}{dr^2}T_N-\left(\frac{dT_N}{dr}\right)^2\right]
                  -\frac{dT_N}{dr}T_N+(4r+2v+1)T_N^2,  \label{rec:T}  
\end{equation}
with $T_0=T_1=1$. Moreover, it is shown that 
\begin{equation}
w(r)=\frac{T_{N+1}(r,v-1)T_N(r,v)}{T_{N+1}(r,v)T_N(r,v-1)}
\end{equation}
gives the rational solutions of P$_{\rm III}$ (\ref{P3}) with parameters 
\begin{equation}
\alpha=4(v+N), \quad \beta=4(-v+N). 
\end{equation}
The polynomials $S_N$ are called the Umemura polynomials. 
Of course, the polynomiality of $S_N$ is far from obvious. 

Let $s_Y(t)$, $t=(t_1,t_2,\cdots)$ be the Schur polynomial associated with the partition $Y$. 
By the Jacobi-Trudi formula, 
it is known that $s_Y$, $Y=(i_1,i_2,\cdots i_N)$, $i_1\geq i_2\geq \cdots\geq i_N>0$ 
is expressed by 
\begin{equation}
s_{(i_1,i_2,\cdots,i_N)}(t_1,t_2,\cdots)=
 \left|
  \begin{array}{cccc}
   p_{i_1}     & p_{i_1+1}   & \cdots & p_{i_1+N-1}  \\
   p_{i_2-1}   & p_{i_2}     & \cdots & p_{i_2+N-2}  \\
   \vdots      & \vdots      & \ddots & \vdots       \\
   p_{i_N-N+1} & p_{i_N-N+2} & \cdots & p_{i_N}
  \end{array}
 \right|,
\end{equation}
where $p_k$'s are defined by the generating function, 
\begin{equation}
\sum_{k=0}^{\infty}p_k(t_1,t_2,\cdots)\lambda^k=\exp\left(\sum_{l=1}^{\infty}t_l\lambda^l\right), 
\quad p_k=0\ \mbox{for}\ k<0. \label{gf:Schur}
\end{equation}
Then our main result is stated as follows. 
\begin{thm}\label{main}
For $N \ge 0$, we have 
\begin{equation}
T_{N+1}=(2r)^{-N(N+1)/2}c_N F_N,  \label{TF}
\end{equation}
where 
\begin{equation}
c_N=(2N-1)!!(2N-3)!!\cdots3!!1!!,
\end{equation}
\begin{equation}
F_N=s_{(N,N-1,\cdots,1)}(t_1,t_2,\cdots),
\end{equation}
\begin{equation}
t_1=2r+v+\frac{1}{2},\quad t_k=(-1)^{k+1}\frac{v+\frac{1}{2}}{k},~(k\ge2). \label{spec}
\end{equation}
\end{thm}
This theorem gives an explicit form of the Umemura polynomials $S_N(u,v)$ in terms of the 
Schur polynomials $s_Y(t_1,t_2,\cdots)$ associated with partition $Y=(N,N-1,\cdots,1)$.
This also provides the proof of their polynomiality, as shown below.
\begin{cor}
$S_N(u,v)$ are polynomials in $u$ of degree $N(N-1)/2$.
\end{cor}
\begin{pf}
First we note that $s_{(N,N-1,\cdots,1)}(t_1,t_2,\cdots)$
are homogeneous polynomials in $t_1,t_2,\cdots$ with degree $N(N+1)/2$,
if we define the weight of $t_k$ as $k$. Under the specialization (\ref{spec}), $F_N$ are
polynomials in $r$ and thus $1/u$ of degree $N(N+1)/2$. Since $S_{N+1}={\rm
const.}\times u^{N(N+1)/2}F_N$, the statement follows immediately. \qed
\end{pf}

\section{Proof of the theorem}
Under the specialization (\ref{spec}), the generating function (\ref{gf:Schur}) is reduced to 
\begin{equation}
\sum_{k=0}^{\infty}p_k(x,n)\lambda^k=(1+\lambda)^n \exp(x\lambda), 
\quad p_k=0\ \mbox{for}\ k<0,  \label{gf:Laguerre}
\end{equation}
with 
\begin{equation}
x=2r, \quad n=v+\frac{1}{2}. \label{vt}
\end{equation}
The polynomials $p_k(x,n)$ are essentially the Laguerre polynomials, i.e. 
$p_k(x,n)=L_k^{(n-k)}(-x)$~\cite{Abramowitz}. 

The proof of Theorem \ref{main} is done by a completely parallel method to that in \cite{NY}. 
\begin{prop}\label{prop}
We define $\tau_N=\tau_N(x,n)$ as 
\begin{equation}
\tau_N=
 \left|
  \begin{array}{cccc}
   p_N      & p_{N+1}  & \cdots & p_{2N-1}  \\
   p_{N-2}  & p_{N-1}  & \cdots & p_{2N-3}  \\
   \vdots   & \vdots   & \ddots & \vdots    \\
   p_{-N+2} & p_{-N+3} & \cdots & p_1
  \end{array}
 \right|,\quad N\geq 1,    \label{tau}
\end{equation}
\begin{equation}
 \tau_{-1}=1,\quad \tau_0=1,\label{initial}
\end{equation}
where $p_k$'s are defined by equation (\ref{gf:Laguerre}). 
Then $\tau_N$ satisfy the following recursion relation, 
\begin{equation}
-(2N+1)\tau_{N+1}\tau_{N-1}=
x\left(\tau''_N\tau_N-{\tau'_N}^2\right)+\tau_N\tau'_N-(x+n)\tau_N^2,\ 
N\geq 0,\label{Toda}
\end{equation}
where $\displaystyle '=\frac{d}{dx}$. 
\end{prop}

Theorem \ref{main} is the direct consequence of this proposition, 
since the recursion relation (\ref{Toda}) is reduced to (\ref{rec:T}) 
by equations (\ref{TF}) and (\ref{vt}). 

To prove Proposition \ref{prop}, we prepare the following lemmas. 
\begin{lem}\label{2}
We put $F_N=F_N(x)$ and $G_N=G_N(x)$ as 
\begin{eqnarray}
F_N=
 \left|
  \begin{array}{cccc}
   p_N      & p_{N+1}  & \cdots & p_{2N-1}  \\
   p_{N-2}  & p_{N-1}  & \cdots & p_{2N-3}  \\
   \vdots   & \vdots   & \ddots & \vdots    \\
   p_{-N+2} & p_{-N+3} & \cdots & p_1
  \end{array}
 \right|,\quad N\geq 1, \\
G_N=
 \left|
  \begin{array}{cccc}
   p_{N+2}  & p_{N+3}  & \cdots & p_{2N+1}  \\
   p_{N-2}  & p_{N-1}  & \cdots & p_{2N-3}  \\
   \vdots   & \vdots   & \ddots & \vdots    \\
   p_{-N+2} & p_{-N+3} & \cdots & p_1
  \end{array}
 \right|, \quad N\geq 1,
\end{eqnarray}
\begin{equation}
 F_0=1.
\end{equation}
Then we have 
\begin{equation}
F_{N+1}F_{N-1}=G'_NF_N-G_NF'_N,\quad N\geq 1.   \label{toJacobi}
\end{equation}
\end{lem}
\begin{pf}
Since it is easy to see that 
\begin{equation}
p'_k(x,n)=p_{k-1}(x,n), 
\end{equation}
one can rewrite $F_N$ and $G_N$ as 
\begin{equation}
F_N=
 \left|
  \begin{array}{cccc}
   p_{2N-1}^{(N-1)} & \cdots & p_{2N-1}^{(1)} & p_{2N-1}  \\
   p_{2N-3}^{(N-1)} & \cdots & p_{2N-3}^{(1)} & p_{2N-3}  \\
   \vdots           & \vdots & \ddots         & \vdots    \\
   p_1^{(N-1)}      & \cdots & p_1^{(1)}      & p_1
  \end{array}
 \right|,
G_N=
 \left|
  \begin{array}{cccc}
   p_{2N+1}^{(N-1)} & \cdots & p_{2N+1}^{(1)} & p_{2N+1}  \\
   p_{2N-3}^{(N-1)} & \cdots & p_{2N-3}^{(1)} & p_{2N-3}  \\
   \vdots           & \vdots & \ddots         & \vdots    \\
   p_1^{(N-1)}      & \cdots & p_1^{(1)}      & p_1
  \end{array}
 \right|,
\end{equation}
where $\displaystyle p_k^{(m)}=\frac{d^m}{dx^m}p_k$. 
Putting $D=F_{N+1}$, equation (\ref{toJacobi}) is reduced to Jacobi identity, 
\begin{equation}
D{\left[{{1~2}\atop{1~2}}\right]} D=
D{\left[{{1}\atop{1}}\right]}D{\left[{{2}\atop{2}}\right]}-
D{\left[{{1}\atop{2}}\right]}D{\left[{{2}\atop{1}}\right]}, 
\end{equation}
where the minor $\displaystyle D{\left[{{i_1~i_2~\cdots} \atop {j_1~j_2~\cdots}}\right]}$ 
is defined by deleting the $i_k$-th rows and the $j_k$-th columns from the determinant $D$. 
\qed
\end{pf}
\begin{lem}\label{3}
For $N \ge 1$, 
\begin{eqnarray}
&&(2N+1)s_{(N+2,N-1,\cdots,1)}(t_1,t_2,\cdots) \nonumber \\ 
&&=\left[\frac{t_1^2}{2}+(2N+1)t_2
+\sum_{m=1}^{\infty}(m+2)t_{m+2}\frac{\partial}{\partial t_m}\right]
s_{(N,N-1,\cdots,1)}(t_1,t_2,\cdots). 
\end{eqnarray}
\end{lem}

Since the proof of this lemma is given in \cite{NY}, we omit the detail. 

\begin{lem}\label{4}
For $N \ge 1$, we have 
\begin{equation}
(2N+1)G_N=\left[\frac{x^2}{2}+nx+\frac{(n-N)(n-N-1)}{2}-x\frac{d}{dx}\right]F_N. 
\end{equation}
\end{lem}
\begin{pf}
Under the specialization (\ref{spec}) and (\ref{vt}), we have 
\begin{eqnarray}
\sum_{m=1}^{\infty}(m+2)t_{m+2}\frac{\partial}{\partial t_m}
&=&\sum_{m=1}^{\infty}\left[(m+2)t_{m+2}-mt_m\right]\frac{\partial}{\partial t_m}+E\nonumber\\
&=&-x\frac{d}{dx}+E.  
\end{eqnarray}
Here the Euler operator 
\begin{equation}
E=\sum_{m=1}^{\infty}mt_m\frac{\partial}{\partial t_m}
\end{equation}
acts on $s_{(N,N-1,\cdots,1)}(t)$ as $E=N(N+1)/2$, 
since $s_{(N,N-1,\cdots,1)}(t)$ is a homogeneous polynomial of degree $N(N+1)/2$ 
in $t_1,t_2,\cdots$ 
if we define the weight of $t_m$ as $m$. 
Then Lemma \ref{4} follows immediately from Lemma \ref{3}. 
\qed
\end{pf}
Finally, the proof of Proposition \ref{prop} with $N\geq 1$ is obvious
from  Lemmas \ref{2} and \ref{4}. Moreover, the case of $N=0$ is easily
checked by noticing that $\tau_1=p_1=x+n$ and equation (\ref{initial}).
This completes the proof of Theorem \ref{main}.

\section{Degeneration to the rational solutions of P$_{\rm II}$}
In the previous sections, we have shown that 
\begin{equation}
w(x)=\frac{\tau_{N+1}(x,v-1)\tau_N(x,v)}{\tau_{N+1}(x,v)\tau_N(x,v-1)},  \label{dvt}
\end{equation}
where $\tau_N$ is given by equations (\ref{tau}) and (\ref{initial}), satisfies P$_{\rm III}$ 
\begin{equation}
\frac{d^2w}{dx^2}=\frac{1}{w}\left(\frac{dw}{dx}\right)^2-\frac{1}{x}\frac{dw}{dx}
                  +\frac{2}{x}\left(\alpha w^2+\beta\right)+w^3-\frac{1}{w},   \label{P3'}
\end{equation}
with
\begin{equation}
 \alpha = v+N+1,\quad \beta=-v+N+1.
\end{equation}
In this section, we consider the coalescence procedure from P$_{\rm III}$ to P$_{\rm II}$ 
on the level of the rational solutions. 

In \cite{p2:rational}, it is shown that the rational solutions of P$_{\rm II}$ 
are expressed as follows. 
\begin{prop}
Let $q_k(z)$ be a sequence of polynomials defined by the generating function, 
\begin{equation}
\sum_{k=0}^\infty q_k(z)~\xi^k=\exp\left(z\xi+\frac{\xi^3}{3}\right),
\quad q_k(z)=0\ {\rm for}\ k<0\ ,   \label{gf:q}
\end{equation}
and let $\sigma_N$ be an determinant given by 
\begin{equation}
\sigma_N=
 \left|
  \begin{array}{cccc}
   q_N      & q_{N+1}  & \cdots & q_{2N-1} \\
   q_{N-2}  & q_{N-1}  & \cdots & q_{2N-3} \\
   \vdots   & \vdots   & \ddots & \vdots   \\
   q_{-N+2} & q_{-N+3} & \cdots & q_1
  \end{array}
 \right|,\quad \sigma_0=1.
\end{equation}
Then 
\begin{equation}
\tilde{w}=\frac{d}{dz}\log\frac{\sigma_{N+1}}{\sigma_N}  \label{dvt:P2}
\end{equation}
gives the rational solutions of P$_{\rm II}$, 
\begin{equation}
\frac{d^2\tilde{w}}{dz^2}=2\tilde{w}^3-4z\tilde{w}+4(N+1). \label{P2}
\end{equation}
\end{prop}

It is easy to see that by putting~\cite{Ince,Tami} 
\begin{equation}
x=\epsilon^{-3}(1-\epsilon^2z), \quad v=-\epsilon^{-3}, \quad w=1+\epsilon \tilde{w}, \label{vP2}
\end{equation}
P$_{\rm III}$ (\ref{P3'}) is reduced to P$_{\rm II}$ (\ref{P2}) in the limit of $\epsilon \to 0$. 

Next, we consider the degeneration of $\tau$-function. 
Let $\bar{p}_k(x,v)$ be a sequence of polynomials defined by the generating function, 
\begin{equation}
\sum_{k=0}^{\infty}\bar{p}_k(x,v)\lambda^k
=(1+\lambda)^{v+\frac{1}{2}} \exp(x\lambda+\frac{v}{2}\lambda^2), 
\quad p_k=0\ \mbox{for}\ k<0,  \label{gf:p-bar}
\end{equation}
then $\tau$-function (\ref{tau}) is rewritten as 
\begin{equation}
\tau_N(x,v)=
 \left|
  \begin{array}{cccc}
   \bar{p}_N      & \bar{p}_{N+1}  & \cdots & \bar{p}_{2N-1}  \\
   \bar{p}_{N-2}  & \bar{p}_{N-1}  & \cdots & \bar{p}_{2N-3}  \\
   \vdots         & \vdots         & \ddots & \vdots          \\
   \bar{p}_{-N+2} & \bar{p}_{-N+3} & \cdots & \bar{p}_1
  \end{array}
 \right|, 
\end{equation}
since $\bar{p}_k(x)$ is a linear combination of $p_j(x), j=k,k-2,k-4,\cdots$. 
Putting $\lambda = -\epsilon \xi$ and $\bar{q}_k=(-\epsilon)^k \bar{p}_k$ 
in equation (\ref{gf:p-bar}) and choosing the parameters as equation (\ref{vP2}), we have 
\begin{equation}
\sum_{k=0}^{\infty}\bar{q}_k(z,v)\xi^k = 
\exp\left[\left(z\xi+\frac{\xi^3}{3}\right)+\epsilon\left(-\frac{\xi}{2}+\frac{\xi^4}{4}\right)
+O(\epsilon^2)\right].  \label{gf:q-bar}
\end{equation}
Since we obtain, from equations (\ref{gf:q}) and (\ref{gf:q-bar}), 
\begin{equation}
\bar{q}_k(z,v)
=q_k(z)+\epsilon\left[-\frac{1}{2}q_{k-1}(z)+\frac{1}{4}q_{k-4}(z)\right]+O(\epsilon^2), 
\label{q:v}
\end{equation}
$\bar{q}_k$ is reduced to $q_k$. 
Thus, we find that $\tau_N$ degenerates to $(-\epsilon)^{-N(N+1)/2}\sigma_N$ 
in the limit of $\epsilon \to 0$. 
We notice that the overall factor $(-\epsilon)^{-N(N+1)/2}$ on $\sigma_N$ has 
no effect on the solutions $\tilde{w}$. 

Finally, we show that the solutions (\ref{dvt}) degenerate to (\ref{dvt:P2}). 
Similarly to the above discussion, we have 
\begin{equation}
\bar{q}_k(z,v-1)
=q_k(z)+\epsilon\left[\frac{1}{2}q_{k-1}(z)+\frac{1}{4}q_{k-4}(z)\right]+O(\epsilon^2). 
\label{q:v-1}
\end{equation}
Since it is easy to see that 
\begin{equation}
\frac{d}{dz}q_k=q_{k-1}, 
\end{equation}
we obtain the following relations, 
\begin{equation}
 \left.
  \begin{array}{@{\,}ll}
   \displaystyle
   \tau_N(v) \simeq 
    \sigma_N-\frac{\epsilon}{2}\frac{d}{dz}\sigma_N+\epsilon\kappa_N+O(\epsilon^2),  \\
   \displaystyle
   \tau_N(v-1) \simeq 
    \sigma_N+\frac{\epsilon}{2}\frac{d}{dz}\sigma_N+\epsilon\kappa_N+O(\epsilon^2),
  \end{array}
 \right.
\end{equation}
where $\kappa_N$ denotes the contribution from the third term of 
equations (\ref{q:v}) and (\ref{q:v-1}). 
Thus, we find that 
\begin{equation}
\frac{\tau_{N+1}(v-1)\tau_N(v)}{\tau_{N+1}(v)\tau_N(v-1)}
=1+\epsilon\frac{d}{dz}\log\frac{\sigma_{N+1}}{\sigma_N}+O(\epsilon^2). 
\end{equation}
Hence the rational solutions of P$_{\rm III}$ are reduced to those of P$_{\rm II}$ 
consistently under the parametrization (\ref{vP2}) in the limit of $\epsilon \to 0$.

\section{Concluding remarks}
In this letter, we have shown that the Umemura polynomials for P$_{\rm III}$ admit 
a determinant expression of Jacobi-Trudi type whose entries are the Laguerre polynomials. 
Moreover we have mentioned that this determinant expression degenerates to that for 
the rational solutions of P$_{\rm II}$ by applying the coalescence procedure. 

It is worth to remark that the Umemura polynomials for P$_{\rm V}$ have a quite similar 
expression, which is also the special case of 
Jacobi-Trudi formula for the 2-reduced Schur polynomials~\cite{NY}. 
The only difference is that we take $p_k$, entries of determinant, as $L_k^{(n-k)}$ 
in the case of P$_{\rm III}$, while $L_k^{(n)}$ in the case of P$_{\rm V}$. 
This strange resemblance should have theoretical explanation. 

It is known that B\"acklund transformation of P$_{\rm III}$ is 
nothing but the alternate discrete Painlev\'e II equation (alt-dP$_{\rm II}$)~\cite{alt-dp2}. 
This suggests that the rational solutions of alt-dP$_{\rm II}$ admit 
the same determinant expression as that for P$_{\rm III}$. 
In this case, the parameter $v$ plays a role of the independent variable. 

Now it has been revealed that special polynomials associated with the rational solutions of 
Painlev\'e equations are special cases of the Schur polynomials, except for P$_{\rm VI}$. 
Since P$_{\rm VI}$ is a ``master'' equation among the Painlev\'e equations, 
it is an important problem to what kind of polynomials the Umemura polynomials 
for P$_{\rm VI}$ are identified. 
Moreover, studying special polynomials of various discrete, $q$-discrete and ultradiscrete 
analogues of the Painlev\'e equations might be an interesting problem.

\ack 
The authors would like to thank Prof. M. Noumi and Prof. Y. Yamada for discussions. 
They also thank Prof. H. Umemura for encouragement. 
One of the authors (K.K.) is supported by the Grant-in-aid for Encouragement of Young Scientists, 
The Ministry of Education, Science and Culture of Japan, No. 09740164.


\end{document}